# Strictly Monotone Brouwer Trees for Well-founded Recursion Over Multiple Arguments

Joseph Eremondi
Laboratory for the Foundations of Computer Science
University of Edinburgh
Scotland, United Kingdom
joey.eremondi@ed.ac.uk

## Abstract

Ordinals can be used to prove the termination of dependently typed programs. Brouwer trees are a particular ordinal notation that make it very easy to assign sizes to higher order data structures. They extend unary natural numbers with a limit constructor, so a function's size can be the least upper bound of the sizes of values from its image. These can then be used to define well-founded recursion: any recursive calls are allowed so long as they are on values whose sizes are strictly smaller than the current size.

Unfortunately, Brouwer trees are not algebraically well-behaved. They can be characterized equationally as a join-semilattice, where the join takes the maximum of two trees. However, this join does not interact well with the successor constructor, so it does not interact properly with the strict ordering used in well-founded recursion.

We present Strictly Monotone Brouwer trees (SMB-trees), a refinement of Brouwer trees that are algebraically well-behaved. SMB-trees are built using functions with the same signatures as Brouwer tree constructors, and they satisfy all Brouwer tree inequalities. However, their join operator distributes over the successor, making them suited for well-founded recursion or equational reasoning.

This paper teaches how, using dependent pairs and careful definitions, an ill-behaved definition can be turned into a well-behaved one. Our approach is axiomatically lightweight: it does not rely on Axiom K, univalence, quotient types, or Higher Inductive Types. We implement a recursively-defined maximum operator for Brouwer trees that matches on successors and handles them specifically. Then, we define SMB-trees as the subset of Brouwer trees for which the recursive maximum computes a least upper bound. Finally, we show that every Brouwer tree can be transformed into a corresponding SMB-tree by infinitely joining it with itself. All definitions and theorems are implemented in Agda.

## 1 Introduction

### 1.1 Recursion and Dependent Types

Dependently typed programming languages bridge the gap between theorem proving and programming. In languages like Agda [Norell 2009], Coq [Bertot and Castéran 2004], Idris [Brady 2021], and Lean [de Moura et al. 2015], one can write programs, specifications, and proofs that programs meet those specifications, all using a unified language.

One challenge in writing dependently typed code is proving termination. Functions in dependently typed languages are typically required to be *total*: they must provably halt in all inputs. This is necessary both to ensure that type checking terminates and to prevent false results from being accidentally proven. Since the halting problem is undecidable, recursively-defined functions must be written in such a way that the type checker can mechanically deduce termination. Some functions only make recursive calls to structurally-smaller arguments, so their termination is apparent to the compiler. However, some functions are not easily expressed using structural recursion. For such functions, the programmer must instead use *well-founded recursion*, showing that there is some ordering, with no infinitely-descending chains, for which each recursive call is strictly smaller according to this ordering. For example, a typical quicksort is not structurally recursive, but can use well-founded recursion on the length of the lists being sorted.

### 1.2 Ordinals

While numeric orderings work for first-order data, they are ill-suited to recursion over higher-order data structures, where some fields contain functions. Instead, one must use *ordinals* to assign a size to such data structures, so that even when a structure represents infinite data, only a finite number of recursive calls are made when traversing it. In classical mathematics, ordinals are totally ordered and straightforward to reason about. They have many different representations, all of which are equivalent. However, in constructive theories, such as those underlying dependently typed languages, there are many representations of ordinals which are not equivalent. Different constructive ordinal notations have different capabilities, each with their own advantages and disadvantages.

### 1.3 Contributions

This work defines *strictly monotone Brouwer Trees*, henceforth SMB-trees, a new presentation of ordinals that hit a sort of sweet-spot for defining functions by well-founded recursion. Specifically, SMB-trees:

- Are strictly ordered by a well-founded relation;
- Have a maximum operator which computes a least-upper bound;
- Are *strictly-monotone* with respect to the maximum: if $a < b$ and $c < d$, then $\max a\ c < \max b\ d$;
- Can compute the limits of arbitrary sequences;
- Are light in axiomatic requirements: they are defined without using axiom K, univalence, quotient types, or higher inductive types.

The novel insight behind our contribution is that there is a subset of Brouwer trees which behave in the way we want. Specifically, the ability of Brouwer trees to take the limit of a sequence allows us to apply operations to an ordinal an infinite number of times, exposing properties that do not hold for finite applications but do hold in the limit.

### 1.4 Uses for SMB-trees

**Well-founded Recursion**   Having a maximum operator for ordinals is particularly useful when traversing over multiple higher order data structures in parallel, where neither argument takes priority over the other. In such a case, a lexicographic ordering cannot be used.

As an example, consider a unification algorithm that merges two higher order data structures, such as a unifier for a strongly typed encoding of dependent types, and suppose that $\alpha$-renaming or some other restriction prevents structural recursion from being used.

To solve a unification problem $\Sigma(x : A).\ B = \Sigma(x : C).\ D$ we must recursively solve $A = C$ and $\forall x.\ B[x] = D[x]$. However, the types of the variables in the latter equation are different. So after computing the unification of $A$ and $C$, we may need to traverse $B$ and $D$ and convert terms from type $A$ or $C$ to their unification. If such a conversion is defined mutually with unification, then it must work on a pair of types strictly smaller than $\Sigma(x : A).\ B, \Sigma(x : C).\ D$.

To assign sizes to such a procedure, we need a few features. First, we need a maximum operator, so that we can bound the size of unifying $A$ and $C$ by their maximum size. Second, the operator should be strictly monotone, so that the recursive call unifying $A$ and $C$ is on a strictly smaller size. Third, the maximum should be commutative: we need the size of the nested pairs $((A, B), (C, D))$ to be the same as $((A, C), (B, D))$, so that a recursive call on arguments whose size is bounded by the maximum of $(A, C)$ will still be strictly smaller than the initial size of $((A, B), (C, D))$. One such call would be the procedure converting from type $A$ to the solution of $A = C$. Lexicographic orderings lack this commutativity, and are too restrictive for situations such as this.

This style of induction was used to prove termination in a syntactic model of gradual dependent types [Eremondi 2023b]. There, Brouwer trees were used to establish termination of recursive procedures that combined the type information in two imprecise types. The decreasing metric was the maximum size of the codes for the types being combined. Brouwer trees' arbitrary limits were used to assign sizes to dependent function and product types, and the strict monotonicity of the maximum operator was essential for proving that recursive calls were on strictly smaller arguments.

We want to enable the programmer to specify complex relationships between the sizes of multiple arguments and to deduce facts about those sizes in a principled way.

**Syntactic Models and Sized Types**   An alternate view of our contribution is as a tool for modelling sized types [Hughes et al. 1996]. The implementation of sized types in Agda has been shown to be unsound [Agda-Developers 2017], due to the interaction between propositional equality and the top size $\infty$ satisfying $\infty < \infty$. Chan [2022] defines a dependently typed language with sized types that does not have a top size, proving it consistent using a syntactic model based on Brouwer trees.

SMB-trees provide the capability to extend existing syntactic models to sized types with a maximum operator. This brings the capability of consistent sized types closer to feature parity with Agda, which has a maximum operator for its sizes, while still maintaining logical consistency.

**Algebraic Reasoning**   Another advantage of SMB-trees is that they allow Brouwer trees to be understood using algebraic tools. In algebraic terminology, SMB-trees satisfy the following algebraic laws, up to the equivalence relation defined by $s \approx t := s \leq t \leq s$

- Join-semlattice: the binary max is associative, commutative, and idempotent;
- Bounded: there is a least tree $Z$ such that $\max t\ Z \approx t$;
- Inflationary endomorphism: there is a successor operator $\uparrow$ such that $\max (\uparrow t)\ t \approx \uparrow t$ and
  $\uparrow(\max s\ t) \approx \max(\uparrow s)\ (\uparrow t)$;

Bezem and Coquand [2022] describe a polynomial time algorithm for solving equations in such an algebra, and describe its usefulness for solving constraints involving universe levels in dependent type checking. While equations involving limits of infinite sequences are undecidable, the inflationary laws could be used to automatically discharge some equations involving sizes. This algebraic presentation is particularly amenable to solving equations using free extensions of algebras [Allais et al. 2023; Corbyn 2021].

### 1.5 Implementation

We have implemented SMB-trees in Agda 2.6.4 with std-lib 1.7.3. Our library specifically avoids Agda-specific features such as cubical type theory or Axiom K, so we expect that it can be easily ported to other proof assistants.

This paper is written as a literate Agda document, and the definitions given in the paper are valid Agda code. For several definitions, only the type is presented, with the body omitted due to space restrictions. The full implementation is open source [Eremondi 2023a].

## 2 Brouwer Trees: An Introduction

Brouwer trees [Church 1938; Kleene 1938] are a simple but elegant tool for proving termination of higher-order procedures. Traditionally, they are defined as follows:

```
data SmallTree : Set where
  Z : SmallTree
  ↑ : SmallTree → SmallTree
  Lim : (ℕ → SmallTree) → SmallTree
```

A Brouwer tree is either zero, the successor of another Brouwer tree, or the limit of a countable sequence of Brouwer trees. However, these are quite weak, in that they can only take the limit of countable sequences. To represent the limits of uncountable sequences, we can parameterize our definition over some Universe à la Tarski:

```
module Brouwer {ℓ}
  (ℂ : Set ℓ)
  (El : ℂ → Set ℓ)
  (ℂℕ : ℂ) (ℂℕIso : Iso (El ℂℕ) ℕ ) where
```

Our module is parameterized over a universe level, a type $\mathbb{C}$ of *codes*, and an "elements-of" interpretation function El, which computes the type represented by each code. We require a code whose interpretation is isomorphic to the natural numbers, as this is essential to our construction in Section 4.1. This also ensures that our trees are at least as powerful as SmallTree. Increasingly larger ordinals can be obtained by setting $\mathbb{C} := \text{Set } \ell$ and El := id for increasing $\ell$. However, by defining an inductive-recursive universe, one can still capture limits over some non-countable types, since Tree is in Set 0 whenever $\mathbb{C}$ is.

Given our universe of codes, we generalize limits to any function whose domain is the interpretation of some code.

```
data Tree : Set ℓ where
  Z : Tree
  ↑ : Tree → Tree
  Lim : (c : ℂ) → (f : El c → Tree) → Tree
```

The small limit constructor can be recovered from the natural-number code

```
ℕLim : (ℕ → Tree) → Tree
ℕLim f = Lim ℂℕ (λ cn → f (Iso.fun ℂℕIso cn))
```

Brouwer trees are the quintessential example of a higher-order inductive type[1]: each tree is built using smaller trees or functions producing smaller trees, which is essentially a way of storing a possibly infinite number of smaller trees.

### 2.1 Ordering Trees

Our ultimate goal is to have a well-founded ordering[2], so we define a relation to order Brouwer trees.

```
data _≤_ : Tree → Tree → Set ℓ where
  ≤-Z : ∀ {t} → Z ≤ t
  ≤-sucMono : ∀ {t₁ t₂}
    → t₁ ≤ t₂
    → ↑ t₁ ≤ ↑ t₂
  ≤-cocone : ∀ {t} {c : ℂ} (f : El c → Tree) (k : El c)
    → t ≤ f k
    → t ≤ Lim c f
  ≤-limiting : ∀ {t} {c : ℂ}
    → (f : El c → Tree)
    → (∀ k → f k ≤ t)
    → Lim c f ≤ t
```

There are four constructors. First, zero is less than any other tree. Second, the successor operator is monotone: if $t_1 ≤ t_2$, then $↑t_1 ≤ ↑t_2$. Finally, there are two constructors which establish that Lim $c$ $f$ denotes the least upper bound of the image of $f$. First ≤-cocone establishes that $f$ $x$ ≤ $Lim\ c\ f$, i.e., it is an upper bound on the image of $f$. Second, ≤-limiting establishes that if a value is an upper bound on the image of $f$, then Lim $c$ $f$ is less than that value, i.e. it is the least of all upper bounds. The constructor names and types are adapted from Kraus et al. [2023], although we change the definition of ≤-cocone slightly so that we do not need a separate constructor for transitivity.

This relation is reflexive:

```
≤-refl : ∀ t → t ≤ t
≤-refl Z = ≤-Z
≤-refl (↑ t) = ≤-sucMono (≤-refl t)
≤-refl (Lim c f)
  = ≤-limiting f (λ k → ≤-cocone f k (≤-refl (f k)))
```

Crucially, it is also transitive, making the relation a preorder.

---

[1]Not to be confused with Higher Inductive Types (HITs) from Homotopy Type Theory [Univalent Foundations Program 2013]
[2]Technically, this is a well-founded quasi-ordering because there are pairs of trees which are related by both ≤ and ≥, but which are not propositionally equal.

```
≤-trans : ∀ {t₁ t₂ t₃} → t₁ ≤ t₂ → t₂ ≤ t₃ → t₁ ≤ t₃
≤-trans ≤-Z p23 = ≤-Z
≤-trans (≤-sucMono p12) (≤-sucMono p23)
    = ≤-sucMono (≤-trans p12 p23)
≤-trans p12 (≤-cocone f k p23)
    = ≤-cocone f k (≤-trans p12 p23)
≤-trans (≤-limiting f x) p23
    = ≤-limiting f (λ k → ≤-trans (x k) p23)
≤-trans (≤-cocone f k p12) (≤-limiting .f x)
    = ≤-trans p12 (x k)
```

We create an infix version of transitivity for more readable construction of proofs:

```
_≤⨾_ : ∀ {t₁ t₂ t₃} → t₁ ≤ t₂ → t₂ ≤ t₃ → t₁ ≤ t₃
lt1 ≤⨾ lt2 = ≤-trans lt1 lt2
```

A useful property is that limits of sequences are related if the sequences are related element-wise:

```
extLim : ∀ {c : ℂ}
    → (f₁ f₂ : El c → Tree)
    → (∀ k → f₁ k ≤ f₂ k)
    → Lim c f₁ ≤ Lim c f₂
extLim {c = c} f₁ f₂ all
    = ≤-limiting f₁ (λ k → ≤-cocone f₂ k (all k))
```

**Strict Ordering** We can define a strictly-less-than relation in terms of our less-than relation and the successor constructor:

```
_<_ : Tree → Tree → Set ℓ
t₁ < t₂ = ↑ t₁ ≤ t₂
```

That is, $t_1$ is strictly smaller than $t_2$ if the tree one-size larger than $t_1$ is as small as $t_2$. The fact that ↑$t$ is always strictly larger than $t$ is a key property of ordinals. Adding one element to a countably-infinite set does not change its cardinality, but taking the successor of an infinite ordinal produces something larger, which is why they are useful for assigning sizes to infinite data.

This relation has the properties one expects of a strictly-less-than relation: it is a transitive sub-relation of the less-than relation, every tree is strictly less than its successor, and no tree is strictly smaller than zero.

```
≤↑t : ∀ t → t ≤ ↑ t
≤↑t Z = ≤-Z
≤↑t (↑ t) = ≤-sucMono (≤↑t t)
≤↑t (Lim c f)
    = ≤-limiting f λ k →
        (≤↑t (f k))
        ≤⨾ (≤-sucMono (≤-cocone f k (≤-refl (f k))))

<-in-≤ : ∀ {x y} → x < y → x ≤ y

<∘≤-in-< : ∀ {x y z} → x < y → y ≤ z → x < z
```

```
≤∘<-in-< : ∀ {x y z} → x ≤ y → y < z → x < z

¬<Z : ∀ t → ¬(t < Z)
```

### 2.2 Well-founded Induction

Here we recall the definition of a constructive well-founded relation. An element is said to be accessible if all strictly smaller elements are accessible. A relation is then well-founded if all elements are accessible. This is formulated as follows:

```
data Acc {A : Set a}
    (_<_ : A → A → Set ℓ)
    (x : A)
    : Set (a ⊔ ℓ) where
  acc : (rs : ∀ y → y < x → Acc _<_ y) → Acc _<_ x

WellFounded : (A → A → Set ℓ) → Set _
WellFounded _<_ = ∀ x → Acc _<_ x
```

That is, an element of a type is accessible for a relation if all strictly smaller elements of it are also accessible. A relation is well-founded if all values are accessible with respect to that relation. This can then be used to define induction with arbitrary recursive calls on smaller values:

```
wfRec : (P : A → Set ℓ)
    → (∀ x → (∀ y → y < x → P y) → P x)
    → ∀ x → P x
```

The wfRec function is defined using structural recursion on an argument of type Acc, so the type checker accepts it. Well-founded induction computes a fixed point of the function, meaning that the particular proof that the strict order holds is irrelevant:

```
unfold-wfRec : ∀ {x}
    → wfRec P f x ≡ f x (λ y _ → wfRec P f y)
```

Following the construction of Kraus et al. [2023], we can show that the strict ordering on Brouwer trees is well-founded. First, we prove a lemma: if a value is accessible, then all (not necessarily strictly) smaller terms are also accessible.

```
smaller-accessible : (x : Tree)
    → Acc _<_ x → ∀ y → y ≤ x → Acc _<_ y
smaller-accessible x (acc r) y x≤y
    = acc (λ y' y'<y → r y' (<∘≤-in-< y'<y x≤y))
```

Then structural induction shows that all terms are accessible. The key observations are that zero is trivially accessible, since no trees are strictly smaller than it, and that the only way to derive ↑$t$ ≤ (Lim $c$ $f$) is with ≤-cocone, yielding a concrete $k$ for which ↑ $t$ ≤ $f$ $k$, on which we recur.

```
ordWF : WellFounded _<_
ordWF Z = acc λ _ ()
ordWF (↑ x)
    = acc (λ { y (≤-sucMono y≤x)
```

```
      → smaller-accessible x (ordWF x) y y≤x})
ordWF (Lim c f) = acc wfLim
  where
    wfLim : (y : Tree) → (y < Lim c f) → Acc _<_ y
    wfLim y (≤-cocone .f k y<fk)
      = smaller-accessible (f k)
          (ordWF (f k)) y (<-in-≤ y<fk)
```

This lets us use Brouwer trees as the decreasing metric for well-founded recursion. However, the wfRec function only worked with one argument. To handle recursion with more than one argument, we need a way to combine ordinals.

## 3  First Attempts at a Join

One way to do induction over multiple arguments is to do well-founded induction over the maximum of the sizes of those arguments. Doing this requires a maximum function, or in semilattice terminology, a join operator.

Here we present two faulty implementations of a join operator for Brouwer trees. The first uses limits to define the join, but does not satisfy strict monotonicity. The second is defined inductively. Its satisfies strict monotonicity, but fails to be the least of all upper bounds, and requires us to assume that limits are only taken over non-empty types. In Section 4, we define SMB-trees: a refinement of Brouwer trees with the benefits of both versions of the maximum.

### 3.1  Limit-based Maximum

Since the limit constructor finds the least upper bound of the image of a function, it should be possible to define the maximum of two trees as a special case of general limits. Indeed, we can compute the maximum of $t_1$ and $t_2$ as the limit of the function that produces $t_1$ when given 0 and $t_2$ otherwise.

```
limMax : Tree → Tree → Tree
limMax t₁ t₂ = ℕLim λ n → if0 n t₁ t₂
```

This version of the maximum has the properties we want from a maximum function: it is an upper bound on its arguments, and it is idempotent.

```
limMax≤L : ∀ {t₁ t₂} → t₁ ≤ limMax t₁ t₂
limMax≤L {t₁} {t₂}
  = ≤-cocone _ (Iso.inv CℕIso 0)
      (subst
         (λ x → t₁ ≤ if0 x t₁ t₂)
         (sym (Iso.rightInv CℕIso 0))
         (≤-refl t₁))

limMax≤R : ∀ {t₁ t₂} → t₂ ≤ limMax t₁ t₂

limMaxIdem : ∀ {t} → limMax t t ≤ t
limMaxIdem {t} = ≤-limiting _ helper
  where
```

```
    helper : ∀ k → if0 (Iso.fun CℕIso k) t t ≤ t
    helper k with Iso.fun CℕIso k
    ... | zero = ≤-refl t
    ... | suc n = ≤-refl t
```

From these properties, we can compute several other useful properties: monotonicity, commutativity, and that it is in fact the least of all upper bounds.

```
limMaxMono : ∀ {t₁ t₂ t₁' t₂'}
  → t₁ ≤ t₁' → t₂ ≤ t₂'
  → limMax t₁ t₂ ≤ limMax t₁' t₂'

limMaxCommut : ∀ {t₁ t₂} → limMax t₁ t₂ ≤ limMax t₂ t₁

limMaxLUB : ∀ {t₁ t₂ t} → t₁ ≤ t → t₂ ≤ t → limMax t₁ t₂ ≤ t
```

This version of the maximum is a least upper bound: by definition Lim denotes the least upper bound of a function's image, and limMax is simply Lim applied to a function whose image has (at most) two elements.

**Limitation: Strict Monotonicity**  The one crucial property that this formulation lacks is that it is not strictly monotone: we cannot deduce max $t_1$ $t_2$ < max $t_1'$ $t_2'$ from $t_1 < t_1'$ and $t_2 < t_2'$. This is because the only way to construct a proof that ↑$t$ ≤ Lim c f is using the ≤-cocone constructor. So we would need to prove that ↑(max $t_1$ $t_2$) ≤ $t_1'$ or that ↑(max $t_1$ $t_2$) ≤ $t_2'$, which cannot be deduced from the premises alone. What we want is to have ↑max $t_1$ $t_2$ ≤ max(↑$t_1$)(↑$t_2$), so that strict monotonicity is a direct consequence of ordinary monotonicity of the maximum. This is not possible when defining the constructor as a limit.

### 3.2  Recursive Maximum

In our next attempt at defining a maximum operator, we obtain strict monotonicity by making indMax (↑$t_1$)(↑$t_2$) = ↑(indMax $t_1$ $t_2$) hold definitionally. Then, provided indMax is monotone, it will also be strictly monotone.

To do this, we compute the maximum of two trees recursively, pattern matching on the operands. We use a *view* [McBride and McKinna 2004] to identify the cases we are matching on: we are matching on two arguments, each with three possible constructors, but several cases overlap. Using a view type lets us avoid enumerating all nine possibilities when defining the maximum and proving its properties.

To begin, we parameterize our definition over a function yielding some element for any code's type. Having a representative of every code will be useful in computing the maximum of a limit and some other tree, since we do not need to handle the special case where the limit of an empty sequence is zero.

```
module IndMax {ℓ}
  (C : Set ℓ)
  (El : C → Set ℓ)
```

```
(CN : C) (CNIso : Iso (El CN) ℕ )
(default : (c : C) → El c) where
```

We then define our view type:

```
private
  data IndMaxView : Tree → Tree → Set ℓ where
    IndMaxZ-L : ∀ {t} → IndMaxView Z t
    IndMaxZ-R : ∀ {t} → IndMaxView t Z
    IndMaxLim-L : ∀ {t} {c : C} {f : El c → Tree}
      → IndMaxView (Lim c f) t
    IndMaxLim-R : ∀ {t} {c : C} {f : El c → Tree}
      → (∀ {c' : C} {f' : El c' → Tree} → ¬ (t ≡ Lim c' f'))
      → IndMaxView t (Lim c f)
    IndMaxLim-Suc : ∀ {t₁ t₂ } → IndMaxView (↑ t₁) (↑ t₂)
opaque
  indMaxView : ∀ t₁ t₂ → IndMaxView t₁ t₂
```

Our view type has five cases. The first two cases handle when either input is zero, and the next two cases handle when either input is a limit. The final case is when both inputs are successors. The helper indMaxView computes the view for any pair of trees.

The maximum is then defined by pattern matching on the view for its arguments:

```
indMax : Tree → Tree → Tree
indMax' : ∀ {t₁ t₂} → IndMaxView t₁ t₂ → Tree

indMax t₁ t₂ = indMax' (indMaxView t₁ t₂)
indMax' {.Z} {t₂} IndMaxZ-L = t₂
indMax' {t₁} {.Z} IndMaxZ-R = t₁
indMax' {Lim c f} {t₂} IndMaxLim-L
  = Lim c λ x → indMax (f x) t₂
indMax' {t₁} {Lim c f} (IndMaxLim-R _)
  = Lim c (λ x → indMax t₁ (f x))
indMax' {↑ t₁} {↑ t₂} IndMaxLim-Suc = ↑ (indMax t₁ t₂)
```

The maximum of zero and $t$ is always $t$, and the maximum of $t$ and the limit of $f$ is the limit of the function computing the maximum between $t$ and $f\ x$. Finally, the maximum of two successors is the successor of the two maxima, giving the definitional equality we need for strict monotonicity.

This definition only works when limits of all codes are inhabited. The ≤-limiting constructor means that Lim $c\ f$ ≤ Z whenever $El\ c$ is uninhabited. So indMax ↑Z Lim $c\ f$ will not actually be an upper bound for ↑Z if $c$ has no inhabitants. In Section 4.2 we show how to circumvent this restriction.

Under the assumption that all code are inhabited, we obtain several of our desired properties for a maximum: it is an upper bound, it is monotone and strictly monotone, and it is associative and commutative. The proof bodies are omitted: they are straightforward reasoning by cases, but they are long and tedious.

```
opaque
  unfolding indMax indMax'

  indMax-≤L : ∀ {t₁ t₂} → t₁ ≤ indMax t₁ t₂
  indMax-≤L {t₁} {t₂} with indMaxView t₁ t₂
  ... | IndMaxZ-L = ≤-Z
  ... | IndMaxZ-R = ≤-refl _
  ... | IndMaxLim-L {f = f}
    = extLim f (λ x → indMax (f x) t₂) (λ k → indMax-≤L)
  ... | IndMaxLim-R {f = f} _
    = underLim λ k → indMax-≤L {t₂ = f k}
  ... | IndMaxLim-Suc
    = ≤-sucMono indMax-≤L

  indMax-≤R : ∀ {t₁ t₂} → t₂ ≤ indMax t₁ t₂

  indMax-monoL : ∀ {t₁ t'₁ t₂}
    → t₁ ≤ t'₁ → indMax t₁ t₂ ≤ indMax t'₁ t₂
  indMax-monoR : ∀ {t₁ t₂ t'₂}
    → t₂ ≤ t'₂ → indMax t₁ t₂ ≤ indMax t₁ t'₂

  indMax-mono : ∀ {t₁ t₂ t'₁ t'₂}
    → t₁ ≤ t'₁ → t₂ ≤ t'₂ → indMax t₁ t₂ ≤ indMax t'₁ t'₂

  indMax-strictMono : ∀ {t₁ t₂ t'₁ t'₂}
    → t₁ < t'₁ → t₂ < t'₂ → indMax t₁ t₂ < indMax t'₁ t'₂
  indMax-strictMono lt1 lt2 = indMax-mono lt1 lt2

  indMax-assocL : ∀ t₁ t₂ t₃
    → indMax t₁ (indMax t₂ t₃) ≤ indMax (indMax t₁ t₂) t₃
  indMax-assocR : ∀ t₁ t₂ t₃
    → indMax (indMax t₁ t₂) t₃ ≤ indMax t₁ (indMax t₂ t₃)
  indMax-commut : ∀ t₁ t₂
    → indMax t₁ t₂ ≤ indMax t₂ t₁
```

**Limitation: Idempotence**   The problem with an inductive definition of the maximum is that we cannot prove that it is idempotent. Since indMax is associative and commutative, proving idempotence is equivalent to proving that it computes a true least-upper-bound.

The difficulty lies in showing that indMax (Lim $c\ f$) (Lim $c\ f$) ≤ (Lim $c\ f$). By our definition, indMax (Lim $c\ f$) (Lim $c\ f$) reduces to:

$$(\text{Lim}\ c\ \lambda x \to (\text{Lim}\ c\ (\lambda y \to \text{indMax}\ (f\ x)\ (f\ y)))) \leq \text{Lim}\ c\ f$$

We cannot use ≤-cocone to prove this, since the left-hand side is not necessarily equal to $f\ k$ for any $k : El\ c$. So the only possibility is to use ≤-limiting. Applying it twice, along with a use of commutativity of indMax, we are left with the following goal:

$$\forall x \to \forall y \to (\text{indMax}\ (f\ x)\ (f\ y) \leq \text{Lim}\ c\ f)$$

There is no a priori way to prove this goal without already having a proof that indMax is a least upper bound. But proving that was the whole point of proving idempotence! An inductive hypothesis would give that indMax $(f\ x)\ (f\ x) \le f\ x \le Lim\ c\ f$, but it does not apply when the arguments to indMax are not equal. Because we are working with constructive ordinals, we have no trichotomy property [Kraus et al. 2023], and hence no guarantee that indMax $(f\ x)\ (f\ y)$ will be one of $f\ x$ and $f\ y$.

We now have two competing definitions for the maximum: the limit version, which is not strictly monotone, and the inductive version, which is not actually a least upper bound. In the next section, we describe a large class of trees for which indMax is idempotent, and hence does compute a true upper bound. We then use that in Section 4.2 to create a version of ordinals whose join has the best properties of both limMax and indMax.

## 4 Trees with a Strictly-Monotone Idempotent Join

### 4.1 Well-Behaved Trees

Our first step in defining an ordinal notation with a well-behaved maximum is to identify a class of Brouwer trees which are well-behaved with respect to the inductive maximum. As we saw in the previous section, neither the limit based nor the inductive definition of the maximum was satisfactory.

The solution, it turns out, is more limits: if we indMax a term with itself an infinite number of times, the result will be idempotent with respect to indMax. This is essentially an application of Kleene's Fixed-Point Theorem [Cousot and Cousot 1979], using transfinite iteration to find the solution to $x \approx$ indMax $x\ x$.

First, we define a function to indMax a term with itself $n$ times or a given number $n$:

```
nindMax : Tree → ℕ → Tree
nindMax t ℕ.zero = Z
nindMax t (ℕ.suc n) = indMax (nindMax t n) t
```

To compute a tree equivalent to the infinite chain of applications indMax $t$ (indMax $t$ (indMax $t$ ...)), we take the limit of $n$ applications over all $n$:

```
indMax∞ : Tree → Tree
indMax∞ t = ℕLim (λ n → nindMax t n)
```

This operator has useful basic properties: it is monotone, and it computes an upper bound on is argument.

```
indMax∞-self : ∀ t → t ≤ indMax∞ t

indMax∞-mono : ∀ {t₁ t₂}
  → t₁ ≤ t₂
  → (indMax∞ t₁) ≤ (indMax∞ t₂)
```

However, the most important property that we want from indMax∞ is that indMax is idempotent with respect to it. The first step to showing this is realizing that we can take the maximum of $t$ and indMax∞ $t$, and that we have a tree that is no larger than indMax∞ $t$: because it is already an infinite chain of applications, adding one more makes no difference.

```
indMax-∞lt1 : ∀ t → indMax (indMax∞ t) t ≤ indMax∞ t
indMax-∞lt1 t = ≤-limiting _ λ k → helper (Iso.fun CNIso k)
  where
  helper : ∀ n → indMax (nindMax t n) t ≤ indMax∞ t
  helper n =
    ≤-cocone _ (Iso.inv CNIso (ℕ.suc n))
    (subst (λ sn → nindMax t (ℕ.suc n) ≤ nindMax t sn)
      (sym (Iso.rightInv CNIso (suc n)))
      (≤-refl _))
```

If adding one more indMax $t$ has no effect, then adding $n$ more will also have no effect:

```
indMax-∞ltn : ∀ n t
  → indMax (indMax∞ t) (nindMax t n) ≤ indMax∞ t
indMax-∞ltn ℕ.zero t = indMax-≤Z (indMax∞ t)
indMax-∞ltn (ℕ.suc n) t =
  indMax-monoR
    {t₁ = indMax∞ t} (indMax-commut (nindMax t n) t)
  ≤ ⨾ indMax-assocL (indMax∞ t) t (nindMax t n)
  ≤ ⨾ indMax-monoL (indMax-∞lt1 t)
  ≤ ⨾ indMax-∞ltn n t
```

It remains to show that taking indMax of indMax∞ $t$ with itself does not make it larger. By our inductive definition of indMax, we have that

$$\text{indMax (indMax∞ } t)(\text{indMax∞ } t)$$

is equal to

$$\mathbb{N}\text{Lim } (\lambda n \to \text{indMax (nIndMax } n\ t)\ (\text{indMax∞ } t))$$

Our previous lemma gives that, for any $n$, indMax∞ $t$ is an upper bound for indMax (nIndMax $n\ t$) (indMax∞ $t$)). So ≤-limiting gives that the limit over all $n$ is also bounded by indMax∞ $t$, i.e. Lim constructs the least of all upper bounds. This gives us our key result: up to ≤, indMax is idempotent on values constructed with indMax∞.

```
indMax∞-idem : ∀ t
  → indMax (indMax∞ t) (indMax∞ t) ≤ indMax∞ t
indMax∞-idem t =
  ≤-limiting _ λ k →
    (indMax-commut
      (nindMax t (Iso.fun CNIso k)) (indMax∞ t))
  ≤ ⨾ indMax-∞ltn (Iso.fun CNIso k) t
```

There is one last property to prove that will be useful in the next section: indMax∞ $t$ is a lower bound on $t$, and hence equivalent to it, whenever indMax is idempotent on $t$. If taking indMax of $t$ with itself does not increase it size, doing so $n$ times will not increase it size, so again the result follows from Lim being the least upper bound.

```
indMax∞-≤ : ∀ {t} → indMax t t ≤ t → indMax∞ t ≤ t
indMax∞-≤ lt
  = ≤-limiting _
     λ k → nindMax-≤ (Iso.fun CNIso k) lt
  where
    nindMax-≤ : ∀ {t} n → indMax t t ≤ t → nindMax t n ≤ t
    nindMax-≤ ℕ.zero lt = ≤-Z
    nindMax-≤ {t = t} (ℕ.suc n) lt
      = indMax-monoL (nindMax-≤ n lt)
        ≤⨾ lt
```

An immediate corollary of this is that indMax∞ (indMax∞ $t$) is equivalent to indMax∞ $t$.

### 4.2 Strictly Monotone Brouwer Trees

Now that we have identified a substantial class of well-behaved Brouwer trees, we want to define a new type containing only those trees. In this section, we will define strictly monotone Brouwer trees (SMB-trees), and show how they can be given a similar interface to Brouwer trees.

To begin, we declare a new Agda module, with the same parameters we have been working with thus far: a type of codes, interpretations of those codes into types, and a code whose interpretation is isomorphic to ℕ.

```
module SMBTree {ℓ}
  (ℂ : Set ℓ)
  (El : ℂ → Set ℓ)
  (CN : ℂ) (CNIso : Iso (El CN) ℕ ) where
```

Next we import all of our definitions so far, using the "Brouwer" prefix to distinguish them from the trees and ordering we are about to define. Critically, we do not instantiate these with the same interpretation function. Instead, we interpret each code wrapped in Maybe. Note that if a type $T$ is isomorphic to ℕ, then Maybe $T$ is as well. Wrapping in Maybe ensures that we always take Brouwer limits over non-empty sets, an assumption that was critical for the definitions of Section 3.2. Essentially, we are adding an explicit zero to every sequence whose limit we take, so that the sequences are never empty, but the upper bound does not change. This detail is hidden in the interface for SMB-trees: the assumption of non-emptiness is only used in the Brouwer trees underlying SMB-trees.

```
import Brouwer
  ℂ
  (λ c → Maybe (El c))
  CN (maybeNatIso CNIso)
  as Brouwer
```

**Refining Brouwer Trees**   We define SMB-trees as a dependent record, containing an underlying Brouwer tree, and a proof that indMax is idempotent on this tree.

```
record SMBTree : Set ℓ where
  constructor MkTree
  field
    rawTree : Brouwer.Tree
    isIdem : (indMax rawTree rawTree) Brouwer.≤ rawTree
open SMBTree
```

We can then define so-called "smart-constructors" corresponding to each of the constructors for Brouwer-trees: zero, successor, and limit. Zero and successor directly correspond to the Brouwer tree zero and successor. Their proofs of idempotence are trivial from the properties of Brouwer ≤.

```
opaque
  unfolding indMax

  Z : SMBTree
  Z = MkTree Brouwer.Z Brouwer.≤-Z

  ↑ : SMBTree → SMBTree
  ↑ (MkTree t pf)
    = MkTree (Brouwer.↑ t) (Brouwer.≤-sucMono pf)
```

However, constructing the limit of a sequence of SMB-trees is not so easy. Since we instantiated $El$ to wrap its result in Maybe, we need to handle nothing for each limit, but we can use Z as a default value, since adding it to any sequence does not change the least upper bound. More challenging is how, as we saw in Section 3.2, Brouwer trees do not have indMax (Lim $c$ $f$) (Lim $c$ $f$) ≤ Lim $c$ $f$, so we cannot directly produce a proof of idempotence.

Our key insight is to define limits of SMB-trees using indMax∞ on the underlying trees: for any function producing SMB-trees, we take the limit of the underlying trees, then indMax that result with itself an infinite number of times. The idempotence proof is then the property of indMax∞ that we proved in Section 4.1.

```
Lim : ∀ (c : ℂ) → (f : El c → SMBTree) → SMBTree
Lim c f =
  MkTree
  (indMax∞
    (Brouwer.Lim c
      (maybe′ (λ x → rawTree (f x)) Brouwer.Z)))
  (indMax∞-idem _)
```

**Ordering SMB-trees**   SMB-trees are ordered by the order on their underlying Brouwer trees:

```
record _≤_ (t₁ t₂ : SMBTree) : Set ℓ where
  constructor mk≤
  inductive
```

```
  field
    get≤ : (rawTree t₁) Brouwer.≤ (rawTree t₂)
open _≤_
```

The successor function allows us to define a strict order on SMB-trees.

```
_<_ : SMBTree → SMBTree → Set ℓ
_<_ t₁ t₂ = (↑ t₁) ≤ t₂
```

The next step is to prove that our SMB-tree constructors satisfy the same inequalities as Brouwer trees. Since SMB-trees are ordered by their underlying Brouwer trees, most properties can be directly lifted from Brouwer trees to SMB-trees.

```
opaque
  unfolding Z ↑
  ≤↑ : ∀ t → t ≤ ↑ t
  ≤↑ t = mk≤ (Brouwer.≤↑t _)

  _≤⨾_ : ∀ {t₁ t₂ t₃} → t₁ ≤ t₂ → t₂ ≤ t₃ → t₁ ≤ t₃
  _≤⨾_ (mk≤ lt1) (mk≤ lt2) = mk≤ (Brouwer.≤-trans lt1 lt2)

  ≤-refl : ∀ {t} → t ≤ t
  ≤-refl = mk≤ (Brouwer.≤-refl _)
```

The constructors for ≤ each have a counterpart for SMB-trees. For zero and successor, these are trivially lifted.

```
≤-Z : ∀ {t} → Z ≤ t
≤-Z = mk≤ Brouwer.≤-Z

≤-sucMono : ∀ {t₁ t₂} → t₁ ≤ t₂ → ↑ t₁ ≤ ↑ t₂
≤-sucMono (mk≤ lt) = mk≤ (Brouwer.≤-sucMono lt)
```

The constructors for ordering limits require more attention. To show that an SMB-tree limit is an upper bound, we use the fact that the underlying limit was an upper bound, and the fact that indMax∞ is as large as its argument, since the SMB-tree Lim wraps its result in indMax∞. Note that, since we already have transitivity for our new ≤, we can simply show that f k is less than the limit of f, avoiding the more complicated form of ≤-cocone.

```
≤-limUpperBound : ∀ {c : ℂ} → {f : El c → SMBTree}
  → ∀ k → f k ≤ Lim c f
≤-limUpperBound {c = c} {f = f} k
  = mk≤ (Brouwer.≤-cocone _ (just k) (Brouwer.≤-refl _)
           Brouwer.≤ ⨾ indMax∞-self (Brouwer.Lim c _))
```

Finally, we need to show that the SMB-tree limit is less than all other upper bounds. Suppose t : SMBTree is an upper bound for f, and $t_u$ is the underlying tree for t, and $f_u$ computes the underlying trees for f. Then ≤-limiting gives that the underlying tree for t is an upper bound for the trees underlying the image of f. However, the SMB-tree limit wraps its result in indMax∞, so we need to show that indMax∞ of the limit is also less than t′. The monotonicity of indMax∞ then gives that indMax(Lim c $f_u$) is less than indMax∞ t′. In Section 4.1, we showed that indMax∞ had no effect on Brouwer trees that indMax was idempotent on. This is exactly what the isIdem field of SMB-trees contains! So we have indMax∞ t′ ≤ t′, and transitivity gives our result.

```
≤-limLeast : ∀ {c : ℂ} → {f : El c → SMBTree}
  → {t : SMBTree}
  → (∀ k → f k ≤ t) → Lim c f ≤ t
≤-limLeast {f = f} {t = MkTree t idem} lt
  = mk≤ (
    indMax∞-mono
      (Brouwer.≤-limiting _
        (maybe (λ k → get≤ (lt k)) Brouwer.≤-Z))
    Brouwer.≤ ⨾ (indMax∞-≤ idem) )
```

**The Join for SMB-trees** Our whole reason for defining SMB-trees was to define a well-behaved maximum operator, and we finally have the tools to do so. We can define the join in terms of indMax on the underlying trees. The proof that the indMax is idempotent on the result follows from associativity, commutativity, and monotonicity of indMax.

```
opaque
  unfolding indMax Z ↑ indMaxView
  max : SMBTree → SMBTree → SMBTree
  max t₁ t₂ =
    MkTree
      (indMax (rawTree t₁) (rawTree t₂))
      (indMax-swap4 {t₁ = rawTree t₁} {t₁′ = rawTree t₂}
                    {t₂ = rawTree t₁} {t₂′ = rawTree t₂}
        Brouwer.≤ ⨾ indMax-mono (isIdem t₁) (isIdem t₂))
```

For Brouwer trees, indMax had all the properties we wanted except for idempotence. All of these can be lifted directly to SMB-trees:

```
max-≤L : ∀ {t₁ t₂} → t₁ ≤ max t₁ t₂

max-≤R : ∀ {t₁ t₂} → t₂ ≤ max t₁ t₂

max-mono : ∀ {t₁ t₁′ t₂ t₂′} → t₁ ≤ t₁′ → t₂ ≤ t₂′ →
  max t₁ t₂ ≤ max t₁′ t₂′

max-idem≤ : ∀ {t} → t ≤ max t t

max-commut : ∀ t₁ t₂ → max t₁ t₂ ≤ max t₂ t₁

max-assocL : ∀ t₁ t₂ t₃
  → max t₁ (max t₂ t₃) ≤ max (max t₁ t₂) t₃

max-assocR : ∀ t₁ t₂ t₃
  → max (max t₁ t₂) t₃ ≤ max t₁ (max t₂ t₃)
```

In particular, max is strictly monotone, and distributes over the successor:

max-strictMono : ∀ {$t_1$ $t_1'$ $t_2$ $t_2'$ : SMBTree}
  → $t_1$ < $t_1'$ → $t_2$ < $t_2'$ → max $t_1$ $t_2$ < max $t_1'$ $t_2'$

max-sucMono : ∀ {$t_1$ $t_2$ $t_1'$ $t_2'$ : SMBTree}
  → max $t_1$ $t_2$ ≤ max $t_1'$ $t_2'$ → max $t_1$ $t_2$ < max (↑ $t_1'$) (↑ $t_2'$)

However, because we restricted SMB-trees to only contain Brouwer trees that indMax is idempotent on, we can prove that Max is idempotent for SMB-trees:

max-idem : ∀ {t : SMBTree} → max t t ≤ t
max-idem {t = MkTree t pf} = mk≤ pf

These together are enough to prove that our maximum is the least of all upper bounds.

max-LUB : ∀ {$t_1$ $t_2$ t} → $t_1$ ≤ t → $t_2$ ≤ t → max $t_1$ $t_2$ ≤ t
max-LUB lt1 lt2 = max-mono lt1 lt2 ⨾ max-idem

Perhaps surprisingly, this means that an SMB-tree version of limMax is equivalent to max, since they are both the least upper bound. This in turn means that the limit based maximum is strictly monotone for SMB-trees.

ℕLim : (ℕ → SMBTree) → SMBTree
ℕLim f = Lim Cℕ (λ cn → f (Iso.fun CℕIso cn))

max' : SMBTree → SMBTree → SMBTree
max' $t_1$ $t_2$ = ℕLim (λ n → if0 n $t_1$ $t_2$)

max'-≤L : ∀ {$t_1$ $t_2$} → $t_1$ ≤ max' $t_1$ $t_2$

max'-≤R : ∀ {$t_1$ $t_2$} → $t_2$ ≤ max' $t_1$ $t_2$

max'-LUB : ∀ {$t_1$ $t_2$ t} → $t_1$ ≤ t → $t_2$ ≤ t → max' $t_1$ $t_2$ ≤ t

max≤max' : ∀ {$t_1$ $t_2$} → max $t_1$ $t_2$ ≤ max' $t_1$ $t_2$
max≤max' = max-LUB max'-≤L max'-≤R

max'≤max : ∀ {$t_1$ $t_2$} → max' $t_1$ $t_2$ ≤ max $t_1$ $t_2$
max'≤max = max'-LUB max-≤L max-≤R

***Well-founded Ordering on SMB-trees*** Our motivation for defining SMB-trees was defining well-founded recursion, so the final piece of our definition is a proof that the strict ordering of SMB-trees is well-founded. Intuitively this should hold: there are no infinite descending chains of Brouwer trees, and there are fewer SMB-trees than Brouwer trees, so there can be no infinite descending chains of SMB-trees. The key lemma is that an SMB-tree is accessible if its underlying Brouwer tree is.

sizeWF : WellFounded _<_
sizeWF t = sizeAcc (Brouwer.ordWF (rawTree t))
  where
    sizeAcc : ∀ {t}
      → Acc Brouwer._<_ (rawTree t)
      → Acc _<_ t

sizeAcc {t} (acc x)
  = acc ((λ y lt → sizeAcc (x (rawTree y) (get≤ lt))))

Thus, we have an ordinal type with limits, a strictly monotone join, and well-founded recursion.

## 5 An Algebraic Perspective

As a final contribution, we give an algebraic viewpoint of SMBTrees, in terms of equivalences rather than orderings. There are no new results in this section, but the equational view highlights the ways in which a strictly monotone join is useful in reasoning.

SMB-trees cannot be completely characterized using first order equations, since Lim is an infinitary operator. Nevertheless, we anticipate that many of the equations here could be useful in developing automated rewriting tools or tactics for reasoning about SMBTrees. The constraint of $t_1$ < $t_2$ can be translated into the equation ↑ $t_1$ ∨ $t_2$ ≈ $t_2$, which can then be mechanically simplified according to the equations in the following sections.

### 5.1 Semilattices and Setoids

Unfortunately, SMB-trees are only a preorder, not a partial order. Because we are working in vanilla Agda, we have no function extensionality, so applying Lim to definitionally distinct but extensionally equal functions produces trees that are equivalent but not equal. Postulating that equivalent terms are equal would be inconsistent: inductive and limit-based joins are equivalent for SMB-trees, so ↑ ($t_1$ ∨ $t_2$) is equivalent to limMax (↑ $t_1$) (↑ $t_2$), even though their heads are distinct datatype constructors. Likewise, Lim c f is equivalent to Z any time El c is uninhabited.

As such, we present our equations in the setoid style i.e. up to an equivalence relation, but the results in this section could be adapted to quotient types in a system like Cubical Agda [Vezzosi et al. 2019]. First, we establish that SMB-trees are a bounded join-semilattice.

TreeSemiLat : BoundedJoinSemilattice ℓ ℓ ℓ
Carrier TreeSemiLat = SMBTree
_≈_ TreeSemiLat $t_1$ $t_2$ = $t_1$ SMBTree.≤ $t_2$ × $t_2$ SMBTree.≤ $t_1$
_≤_ TreeSemiLat = SMBTree._≤_
_∨_ TreeSemiLat = SMBTree.max
⊥ TreeSemiLat = SMBTree.Z
_ ...

Orderings between trees can then be expressed equationally using the join: $t_1$ is smaller than $t_2$ iff their join is $t_2$.

ord→equiv : ∀ {$t_1$ $t_2$} → $t_1$ ≤ $t_2$ → $t_1$ ∨ $t_2$ ≈ $t_2$
equiv→ord : ∀ {$t_1$ $t_2$} → $t_1$ ∨ $t_2$ ≈ $t_2$ → $t_1$ ≤ $t_2$

This means that our ordering respects equivalence. Additionally, the successor, join and limit constructors are congruences for our equivalence: equivalent inputs produce equivalent outputs. These can be combined with the proof irrelevance of well-founded recursion to rewrite ordering goals according to algebraic laws.

```
≤≈ : ∀ {t₁ t₂ s₁ s₂}
   → s₁ ≤ t₁ → s₁ ≈ s₂ → t₁ ≈ t₂ → s₂ ≤ t₂
<≈ : ∀ {t₁ t₂ s₁ s₂}
   → s₁ < t₁ → s₁ ≈ s₂ → t₁ ≈ t₂ → s₂ < t₂
↑-cong : ∀ {t₁ t₂}
   → t₁ ≈ t₂ → ↑ t₁ ≈ ↑ t₂
Lim-cong : ∀ {c} {f₁ f₂}
   → (∀ x → f₁ x ≈ f₂ x) → Lim c f₁ ≈ Lim c f₂
max-cong : ∀ {s₁ s₂ t₁ t₂}
   → s₁ ≈ s₂ → t₁ ≈ t₂ → s₁ ∨ t₁ ≈ s₂ ∨ t₂
```

This gives us a framework to present the properties of SMB-trees equationally. For instance, the semilattice properties of the join can be given algebraically: a semilattice is a commutative, idempotent semigroup.

```
assoc : ∀ {t₁ t₂ t₃} → t₁ ∨ (t₂ ∨ t₃) ≈ (t₁ ∨ t₂) ∨ t₃
commut : ∀ {t₁ t₂} → t₁ ∨ t₂ ≈ t₂ ∨ t₁
idem : ∀ {t} → t ∨ t ≈ t
```

### 5.2 Successor: The Inflationary Endomorphism

The algebraic version of strict monotonicity is that the successor function is what Bezem and Coquand [2022] call an *inflationary endomorphism*, i.e. a unary operator whose interactions with the join behave like the successor on natural numbers. To our knowledge, SMB-trees are the first ordinal notation in type theory for which the successor is inflationary and arbitrary limits are supported.

There are two laws to inflationary endomorphisms. First, the maximum of ↑ $t$ and $t$ must be ↑ $t$, which captures the idea that $t$ is less that ↑ $t$.

```
↑absorb : ∀ {t} → t ∨ (↑ t) ≈ ↑ t
↑absorb =
   max-mono (≤↑ _) ≤-refl ⨾ max-idem
   , max-≤R
```

Second, the successor must distribute over the join. Recall that this was precisely the condition we used to establish strict monotonicity.

```
↑dist : ∀ {t₁ t₂} → ↑ (t₁ ∨ t₂) ≈ ↑ t₁ ∨ ↑ t₂
↑dist {t₁} {t₂} =
   max-sucMono ≤-refl
   , max-LUB (≤-sucMono max-≤L) (≤-sucMono max-≤R)
```

### 5.3 Characterizing Limits

Finally, we present some equations regarding joins and limits. Since limits are essentially (possibly-)infinitary joins, we write them using $\bigvee$.

```
⋁ : ∀ {c} → (El c → SMBTree) → SMBTree
⋁ f = Lim _ f
```

Limits are an upper bound, so joining any element from a sequence with the limit of that sequence has no effect:

```
⋁-Bound : ∀ {c : 𝒞} {f : El c → SMBTree} {k}
   → f k ∨ (⋁ f) ≈ ⋁ f
⋁-Bound = ord→equiv (≤-limUpperBound _)
```

Moreover, a limit is an actual supremum: the limit of a function is absorbed by any upper bound of the function's image.

```
⋁-Supremum : ∀ {c : 𝒞} {f : El c → SMBTree} {t}
   → (∀ k → f k ∨ t ≈ t) → (⋁ f) ∨ t ≈ t
⋁-Supremum {f = f} lt
   = ord→equiv (≤-limLeast (λ k → equiv→ord (lt k)))
```

The join of a constant, non-empty sequence is the singular element of that sequence:

```
⋁-const : ∀ {c t}
   → El c
   → Lim c (λ _ → t) ≈ t
⋁-const k = ≤-limLeast (λ _ → ≤-refl) , ≤-limUpperBound k
```

The join of an empty sequence is zero:

```
⋁-empty : ∀ {c f}
   → ¬ (El c)
   → Lim c f ≈ Z
⋁-empty empty
   = (≤-limLeast (λ k → contradiction k empty)) , ≤-Z
```

More interesting is the interaction between limits and joins. For two limits over the same index set, the join of those two limits is the same as the limit of joining the sequences together.

```
⋁-distHomo : ∀ {c : 𝒞} {f g : El c → SMBTree}
   → (Lim c f) ∨ (Lim c g) ≈ Lim c (λ x → f x ∨ g x)
⋁-distHomo
   = max-LUB
      (≤-extLim (λ _ → max-≤L))
      (≤-extLim (λ _ → max-≤R))
     , ≤-limLeast
        (λ k → max-mono
           (≤-limUpperBound _)
           (≤-limUpperBound _))
```

We can obtain a more general result, distributing a limit over a join, so long as the limit is over a non-empty sequence.

The join of a non-zero tree with an empty limit will be non-zero, but pushing the join under a limit will produce zero, so the following result only applies to non-empty limits.

```
⋁-distHet : ∀ {c : C} {f : El c → SMBTree} {t}
  → El c
  → ( ⋁ f) ∨ t ≈ ⋁ (λ x → f x ∨ t)
⋁-distHet k
  = max-LUB
      (≤-extLim (λ _ → max-≤L))
      (≤-limUpperBound k ≤ ⨾ (≤-extLim λ _ → max-≤R))
    , ≤-limLeast (λ k → max-monoL (≤-limUpperBound _) )
```

## 6 Discussion
### 6.1 Comparison to Other Ordinal Systems

In the literature, many different variations of ordinals have been presented. To keep our comparison brief, we refer to the work of Kraus et al. [2023]. They give a comprehensive overview of ordinal notation systems in type theory, with a detailed comparison of their comparative properties. They define three different systems: Cantor normal forms that represent ordinals as binary trees, restricted Brouwer trees that represent ordinals as infinitely branching trees, and well-founded types that represent ordinals as types with a certain sort of relation on their elements.

The definitions Kraus et al. give are more restrictive than ours. For example, for Brouwer trees they require that Lim only operate on functions that are strictly increasing, preventing the definition of limMax. These restrictions make their ordinals very well-behaved with respect to propositional equality, so they can examine their mathematical properties. SMB-trees have less rich theory, but the properties they do satisfy are specifically tailored to proving termination of higher-order programs.

***Transitivity, Extensionality and Well-foundedness***  Kraus et al. show three properties for each system they present: transitivity of the ordering, well-foundedness (as in Section 2.2), and *extensionality*, the property that two ordinals are equal iff their sets of smaller terms are equal. They also show a strict version of extensionality for each system: to ordinals are equal iff their sets of strictly smaller terms are equal.

SMB-trees satisfy each of the above properties: the transitivity of ≤ is inherited from Brouwer trees, and we show well-foundedness in Section 2.2. Extensionality for ≤ is trivially true for our setoid version of equivalence. For propositional equality, extensionality cannot be proved without some form of quotient type. We conjecture that the strict order < is not extensional for SMB-trees, since it does not hold for Brouwer trees without quotient types.

Well-founded types lack a basic transitivity property for the strict order: without additional axioms, one cannot conclude $x < z$ from $x \leq y$ and $y < z$. So, though well-founded types have binary and infinitary suprema like SMB-trees, they lack the basic principles for reasoning about strict orders, making them ill-suited for defining recursive procedures.

***Classifiability***  Classifiability is the property that each ordinal is either zero, a successor, or a limit, and that exactly one of those properties holds. Restricted Brouwer trees and Cantor normal forms both satisfy classifiability, but SMB-trees do not. Even our version of Brouwer trees do not have this property: since we allow non-increasing sequences, the limit of the constant-zero sequence is equivalent to zero.

Not having classifiability does negatively affect the decidability properties of SMB-trees. For example, for restricted Brouwer trees, it is decidable whether a tree is infinite or not, but this is not the case for SMB-trees, since some limits are actually finite. However, since SMB-trees are defined specifically around well-founded recursion, losing decidability properties is an acceptable compromise. Additionally, the ability to reason about SMB-trees using the equational style reduces the need to pattern match on them.

***Joins and Suprema***  The main novelty of SMB-trees is the existence of both binary suprema (joins) and infinitary suprema (limits) that interact well with the strict ordering. Cantor normal forms have binary joins and strict monotonicity (as a by-product of decidable ordering), but lack infinitary joins. Well-founded types have binary and infinitary suprema, but without additional axioms even their successor function is not monotone, so strict monotonicity is out of the question. For restricted Brouwer trees, binary joins cannot exist without further axioms. This is an artifact of allowing Lim only on strictly increasing sequences, since it disallows limMax or other similar constructs. So even without strict monotonicity, the capability of SMB-trees exceeds that of restricted Brouwer trees. The cost of this is that SMB-trees fulfill fewer nice properties with respect to propositional equality. Since setoid reasoning is sufficient for well-founded recursion, we find this tradeoff acceptable.

***Conclusion***  Designing an ordinal library is an exercise in compromise, balancing the desired properties with the limitations of decidability and constructive reasoning. With SMB-trees, we have identified a point in the design space well suited to proving termination. The algebraic framework of SMB-trees lays the groundwork for future developments on reasoning mechanically about ordinals. Beyond of our specific use-case, the development of SMB-trees shows that sometimes careful design with dependent types can avoid the need for additional axioms or language features.

## References
Agda-Developers. 2017. Github Issue: Equality is incompatible with sized types. https://github.com/agda/agda/issues/2820.

Guillaume Allais, Edwin Brady, Nathan Corbyn, Ohad Kammar, and Jeremy Yallop. 2023. Frex: dependently-typed algebraic simplification. arXiv:cs.PL/2306.15375

Yves Bertot and Pierre Castéran. 2004. Interactive Theorem Proving and Program Development. Springer-Verlag.

Marc Bezem and Thierry Coquand. 2022. Loop-checking and the uniform word problem for join-semilattices with an inflationary endomorphism. Theoretical Computer Science 913 (2022), 1–7. https://doi.org/10.1016/j.tcs.2022.01.017

Edwin C. Brady. 2021. Idris 2: Quantitative Type Theory in Practice. CoRR abs/2104.00480 (2021). arXiv:2104.00480 https://arxiv.org/abs/2104.00480

Jonathan H.W. Chan. 2022. Sized dependent types via extensional type theory. Master's thesis. University of British Columbia. https://doi.org/10.14288/1.0416401

Alonzo Church. 1938. The constructive second number class. Bull. Amer. Math. Soc. 44, 4 (1938), 224 – 232.

Nathan Corbyn. 2021. Proof Synthesis with Free Extensions in Intensional Type Theory. Technical Report. University of Cambridge. MEng Dissertation.

Patrick Cousot and Radhia Cousot. 1979. Constructive versions of tarski's fixed point theorems. Pacific J. Math. 82, 1 (May 1979), 43–57. https://doi.org/10.2140/pjm.1979.82.43

Leonardo de Moura, Soonho Kong, Jeremy Avigad, Floris van Doorn, and Jakob von Raumer. 2015. The Lean Theorem Prover (System Description). In Automated Deduction - CADE-25, Amy P. Felty and Aart Middeldorp (Eds.). Springer International Publishing, Cham, 378–388.

Joey Eremondi. 2023a. JoeyEremondi/smb-trees: An Agda Library for Strictly Monotone Brouwer Trees. https://doi.org/10.5281/zenodo.10204397

Joseph S. Eremondi. 2023b. On the design of a gradual dependently typed language for programming. Ph.D. Dissertation. University of British Columbia. https://doi.org/10.14288/1.0428823

John Hughes, Lars Pareto, and Amr Sabry. 1996. Proving the Correctness of Reactive Systems Using Sized Types. In Proceedings of the 23rd ACM SIGPLAN-SIGACT Symposium on Principles of Programming Languages (POPL '96). Association for Computing Machinery, New York, NY, USA, 410–423. https://doi.org/10.1145/237721.240882

S. C. Kleene. 1938. On notation for ordinal numbers. The Journal of Symbolic Logic 3, 4 (1938), 150–155. https://doi.org/10.2307/2267778

Nicolai Kraus, Fredrik Nordvall Forsberg, and Chuangjie Xu. 2023. Type-theoretic approaches to ordinals. Theoretical Computer Science 957 (2023), 113843. https://doi.org/10.1016/j.tcs.2023.113843

Conor McBride and James McKinna. 2004. The view from the left. Journal of Functional Programming 14, 1 (2004), 69–111. https://doi.org/10.1017/S0956796803004829

Ulf Norell. 2009. Dependently Typed Programming in Agda. In Proceedings of the 4th International Workshop on Types in Language Design and Implementation (TLDI '09). ACM, New York, NY, USA, 1–2. https://doi.org/10.1145/1481861.1481862

The Univalent Foundations Program. 2013. Homotopy Type Theory: Univalent Foundations of Mathematics. https://homotopytypetheory.org/book, Institute for Advanced Study.

Andrea Vezzosi, Anders Mörtberg, and Andreas Abel. 2019. Cubical Agda: A Dependently Typed Programming Language with Univalence and Higher Inductive Types. Proc. ACM Program. Lang. 3, ICFP, Article 87 (jul 2019), 29 pages. https://doi.org/10.1145/3341691